\documentstyle[12pt]{article}
\input tcilatex
\begin{document}

\begin{center}
{\bf Some aspects of the three dimensional gravity theories with temporal
scalar field.}

\vspace*{1.5cm}

S.M.KOZYREV

Scientific center gravity wave studies ''Dulkyn''.

e-mail: Sergey@tnpko.ru
\end{center}

\vspace*{1.5cm}

\begin{center}
${\bf Abstract}$
\end{center}

The present paper proposes a new explanation for the 3-dimensional Einstein
general theory of relativity which is free of contradictions and consistent
with usual 4-dimensional physics. We discuss the property of the new gravity
theory with temporal scalar field arise in lower-dimensional theories as the
reduction of timelike extra dimension. These ideas we continued by using the
3-dimensional analog of Jordan, Brans-Dicke theory with temporal scalar
field where space and time are treated in different ways.

\section{Introduction}

It is well known since the ancient times that two concepts of space as
absolute or relative are possible; for example Democritos believed in the
existence of both atoms and empty space, while Aristotle with his opinion
that space only epitomizes the place of the objects comes close to a
relational concept of space. Consequently, although Mach was not the first
who formulated mechanics in terms of purely relational quantities his
profound critique of foundations of Newtonian mechanics played a key role in
Einsteins development of general relativity. It has been difficult to
identify spacetime geometries uniquely from statement of Machs principle
since it is difficult to translate it into some precise mathematical
language. Moreover, each statement of Machs principle should be accompanied
by a declaration of the theoretical framework in which it is applied.

Machs principle has also guided us to developments of multidimensional
gravitation theories. The idea of the ordinary spacetime is viewed as a
hypersurface embedded in a higher dimensional space have been considered in
several different contexts, such as strings \cite{c1}, D-branes \cite{c2},
Randall-Sundrum models \cite{c3}, \cite{c4} and non-compactified versions of
Kaluza-Klein theories \cite{c5}. A Campbell-Magaards theorem \cite{c6}, \cite
{c7} and its extension \cite{c8}, \cite{c9} has acquired fundamental
relevance for granting the mathematical consistency of multidimensional
embedding theories and also has been applied to investigate how
lower-dimensional theories could be related to (3+1)-dimensional Einstein
gravity.

The dimensional reduction in order to explain the property of nature is now
widely used in theoretical physics \cite{c10}, \cite{c11}, \cite{c12}. From
the pioneering paper by Jordan \cite{c13} physicists have explored the odd,
however apparently fruitful thought, that Kaluza-Klein theories with special
concern for the idea that the new 5-dimensional metric component, a
spacetime scalar, might play the role of a varying gravitational constant.
However, Jordan and his colleagues took the next step of separating the
scalar field from the original 5-dimensional metric context unified
gravitational-electromagnetic framework.

On the other hand, many physicists have expressed reservations about the
4-dimensional formalism. Though Lorentz recognized Einstein's hypothesis of
geometrization of gravity as one of the basic principles, his enthusiasm for
it gradually waned soon afterwards. In 1910 Lorentz said ''{\it Die
Vorstellung (die auch Redner nur ungen aufgeben w\"urde), dass Raum und Zeit
etwas vollig Verschiedenes seien und dass es eine wahre Zeit gebe (die
Gleichzeitigkeit w\"urde denn unabh\"angig vom Orte bestehen)}'' \cite{c14}.

An investigation into how lower and higher-dimensional theories of gravity
are related to 3-dimensional Universe is therefore well motivated from a
physical point of view. A potential bridge between gravitational theories of
different dimensionality may be found at least two distinct ways by
employing the {\it embedding} or {\it foliation} approaches \cite{c15}. In
the embedding approach, the geometry is determined in terms of quantities
which are defined exclusively in lower-dimensional. In the foliation
approach, the lower-dimensional space-time geometry is determined by
inducing the metric on the leaves, so that in this case the metric tensor
depends on the extra coordinate. According to this world-view, our
perception of normal 3-dimensional space arises because we leave on a domain
wall (time-brane) in a bulk space. This kind of infinitely thin branes
models can, however, be only treated as approximation. One can describe a
thick brane in framework of 4-dimensional general relativity as domain wall
separating two different states{\it \ past} and {\it future} and to an
observer in ordinary space feels the thickness of brane as {\it present}.

This paper details the new approach to the problem of separate the temporal
scalar field from the 4-dimensional metric context of general relativity
which where outlined in previous letter \cite{c17}. Section 2 develops the
formalism related to the correspondence between the ordinary 4-dimensional
gravity with 3-dimensional theories with temporal scalar field. In Section 3
the exact solutions for the static sphere has been obtained. Finally,
Section 4 briefly discussed some aspects of these theories. We choose units
such that G$_N$ = c = 1, and let Latin indices run 1-4 and Greek indices run
1-3.

\section{Dimensional reduction approach}

Embedding theorems of differential geometry provide a natural framework for
relating higher and lower- dimensional theories of gravity. Procedure
introduced by Campbell- Magaards theorems to embed a Riemannian manifold
into Einstein space has been applied to investigate how low- dimensional
theories could be related to (3+1) dimensional Einstein gravity. Such an
embedding has a namber of applications. For example, employing classical
dimensional reduction techniques \cite{c16}, \cite{c11} a 3-dimensional
theory containing temporal scalar field \cite{c17}, \cite{c18} can be
obtained. Such a possibility naturally appears within the frame of
multidimensional models type of Kaluza-Klein theory.

The claim that, in the spirit of Newtonian theory, the space and time are
treated in completely different ways can be obtained by separate the
temporal scalar field from the original 4-dimensional metric context of
general relativity.

A second tenet of this theory is that all classical macroscopic physical
quantities, such as matter density and pressure, could be given a
geometrical interpretation. In this way, it is proposed that the classical
energy-momentum tensor, which enters the right-hand side of the
3-dimensional Einstein equations, can be generated by pure geometrical
means. In other words, it is claimed that geometrical curvature induces
matter in three dimensions, and to an observer in the ordinary space the
extra dimensions (time) appear as the matter source for gravity. However,
induced stress-energy tensor does not in general uniquely determine the
matter content and the interpretation can lead to quite different kinematic
quantities.

One can use the local an isotropic embedding of the 3-dinensional,
Riemannian manifold with line element

\begin{equation}
ds^2=g_{\alpha \beta }(x^\mu )dx^\alpha dx^\beta ,  \label{f1.1}
\end{equation}

in (3+1)-dimensional manifold defined by the metric

\begin{equation}
ds^2=h_{\alpha \beta }dx^\alpha dx^\beta +\phi ^2dt^2,  \label{f1.2}
\end{equation}

where {\it h}$_{\alpha \beta }$ = {\it h}$_{\alpha \beta }$(x$^\mu $,t) and $%
\phi $=$\phi $(x$^\mu $,t) are functions of the (3+1) variables \{x$^\mu $%
,t\} \cite{c8}. The claim that any energy-momentum can be generated by an
embedding mechanism may be translated in geometrical language as saying that
Riemannian 3-dimensional manifold is embeddable into a 4-dimensional Ricci
manifold. In the coordinates (\ref{f1.2}), the field equations take the form

\begin{equation}
\begin{array}{r}
^{(4)}R_{\alpha \beta }=~^{(3)}R_{\alpha \beta }-\frac{\phi _{;\alpha ;\beta
}}\phi -\qquad \qquad \qquad \qquad \qquad \qquad \qquad \qquad \qquad \qquad
\\ 
-\frac 1{2\phi ^2}\left( \frac{\stackrel{\bullet }{\phi }}\phi \stackrel{%
\bullet }{h}_{\alpha \beta }-\stackrel{\bullet \bullet }{h}_{\alpha \beta
}-\frac 12h^{\gamma \delta }\stackrel{\bullet }{h}_{\gamma \delta }\stackrel{%
\bullet }{h}_{\alpha \beta }+h^{\gamma \delta }\stackrel{\bullet }{h}%
_{\alpha \gamma }\stackrel{\bullet }{h}_{\beta \delta }\right) ,
\end{array}
\label{f1.3}
\end{equation}

\begin{equation}
\ ^{(4)}R_{\alpha ^{}y}=\frac \phi 2\nabla _\beta \left( \frac 1\phi
h^{\beta \gamma }\stackrel{\bullet }{h}_{\gamma \alpha }-\frac 1\phi \delta
_\beta ^\alpha h^{\gamma \delta }\stackrel{\bullet }{h}_{\gamma \delta
}\right) ,  \label{f1.4}
\end{equation}

\begin{equation}
\ ^{(4)}R_{yy}=\phi ~\phi _{;\alpha ;\alpha }-\frac 12h^{\gamma \delta }%
\stackrel{\bullet \bullet }{h}_{\gamma \delta }-\frac 12\stackrel{\bullet }{h%
}^{\gamma \delta }\stackrel{\bullet }{h}_{\gamma \delta }+\frac 12h^{\gamma
\delta }\stackrel{\bullet }{h}_{\gamma \delta }\frac{\stackrel{\bullet }{%
\phi }}\phi -\frac 14h^{\gamma \beta }h^{\delta \alpha }\stackrel{\bullet }{h%
}_{\delta \beta }\stackrel{\bullet }{h}_{\gamma \alpha }.  \label{f1.5}
\end{equation}

where overdots denote normal time derivatives. As in classical general
relativity the metric is determined by Einstein's equations.

\begin{equation}
~^{(3)}R_{\alpha \beta }-\frac 12g_{\alpha \beta }^{}~^{(3)}R=-\kappa
~^{(3)}T_{\alpha \beta },  \label{f1.50}
\end{equation}

where $\kappa $ is the Einstein constant and $^{(3)}T_{\alpha \beta }$
energy-momentum tensor, which is a function of all matter fields and metric.
These give the components of the induced energy-momentum tensor since
Einsteins equations (\ref{f1.50}) hold.

\begin{equation}
\begin{array}{r}
^{(3)}T_{\alpha \beta }=\frac{\phi _{;\alpha ;\beta }}\phi -\frac 1{2\phi
^2}\left[ \frac{\stackrel{\bullet }{\phi }}\phi \stackrel{\bullet }{h}%
_{\alpha \beta }-\stackrel{\bullet \bullet }{h}_{\alpha \beta }+h^{\lambda
\mu }\stackrel{\bullet }{h}_{\alpha \lambda ~}\stackrel{\bullet }{h}_{\beta
\mu }-\frac 12h^{\mu \nu }\stackrel{\bullet }{h}_{\mu \nu }~\stackrel{%
\bullet }{h}_{\alpha \beta }\right] + \\ 
+\frac{h_{\alpha \beta }}{8\phi ^2}\left[ \stackrel{\bullet }{h}^{\mu \nu }%
\stackrel{\bullet }{h}_{\mu \nu }+\left( h^{\mu \nu }\stackrel{\bullet }{h}%
_{\mu \nu }\right) ^2\right] .
\end{array}
\label{f1.6}
\end{equation}

It can be shown that the field equations of this theory can be obtained from
the reduced action that are formally similar to the Jordan, Brans-Dicke
action with $\omega $ = 0 \cite{c11}. The Jordan, Brans-Dicke theory of
gravity \cite{c13}, \cite{c19} represents a natural extension of general
relativity, where a nonminimally-coupled scalar field parametrizes the
space-time dependence of Newtons constant. It is expected on the general
ground that dimensional reduction of multi-dimensional theory yields a
(generalized) Jordan, Brans-Dicke theory. It is well known that the Jordan,
Brans-Dicke theory accommodates both Machs principle and Diracs large number
hypothesis \cite{c13}, \cite{c19}. Note that this conclusion depends
crucially on the particular formulation of Machs principle used \cite{c24}.
Newton's gravitational constant G$_N$ replaced by dynamical scalar field
acts as the source of the (local) gravitational coupling with G$_N$ $\sim $ $%
\phi ^{-1}$ and consequently is determined by the total matter in the
universe through auxiliary scalar field equations.

By the way the presented 3-dimensional effective theory is derived by the
integration of the action with respect to the extra dimension as it
customary in Kaluza-Klein compactification. In contrast with standard
Jordan, Brans-Dicke theory this temporal scalar field don't play the role of
a varying gravitational ''constant''. Take it into account one can show that
such 3-dimensional theory introduce a temporal scalar field, which will
(locally and approximately) play the role of Newtonian gravitational
potential. The next step is to suppose that it is possible to find a special
coordinate system, such that we can make the following split of 4
-dimensional metric

\begin{equation}
g_{ab}=\left( 
\begin{array}{cc}
g_{\alpha \beta }+\phi ^2A_\alpha A_\beta  & \phi ^2A_\alpha  \\ 
\phi ^2A_\beta  & \phi ^2
\end{array}
\right)   \label{f1.61}
\end{equation}

where $\phi $ may be regarded as a temporal scalar field, and A$_\alpha $ is
3-dimensional vector, whose role is to be determined later.

\section{Three dimensional gravity theories with temporal scalar field.}

Alternative ways to construct the temporal scalar field theory is to
consider the more general 3-dimensional theory governed by the action where
gravity is coupled to temporary scalar field. From the formal viewpoint,
scalar-tensor models are purely metric theories including a nonminimal
coupling between a temporal scalar field and the curvature scalar R of the
metric g. The most general action can be submitted by allowing $\omega $($%
\phi $) to depend on the scalar field $\phi $ and introducing a cosmological
function $\lambda $($\phi $). The action for these theories in the Jordan
frame is:

\begin{eqnarray}
S=\int \left( f\left( \phi \right) ~^{(3)}R+\frac{\omega \left( \phi \right) 
}\phi g^{\mu \nu }\phi _{,\mu }\phi _{,\nu }-2\phi ^{}\lambda \left( \phi
\right) +L_{matt}\right) \sqrt{\left| g\right| }d^3x~dt,  \label{f1.7}
\end{eqnarray}

The dynamics of the scalar field $\phi $ depends on the functions $f\left(
\phi \right) ,~\omega $($\phi $) and $\lambda $($\phi $). The matter part of
the action $L_{matt}$ depends on the material fields and the metric but does
not involve the scalar field $\phi .$ This ensures a satisfaction of the
''weak equivalence principle'', that is, the statement that the paths of
test particles in gravitational field are independent of their masses. We
may demand that, in a spacetime consistent with Mach's principle, the metric
tensor should be completely determined by the mass distribution in the
universe. Different choices of function $\omega $($\phi $) and $\lambda $($%
\phi $) give different temporal scalar tensor theories. A simple example
elaborates on further possibilities. One can neglect the cosmological
function assuming $\lambda $($\omega $) = 0 and use $f\left( \phi \right)
=\phi ,$ $\omega $($\phi $) = const. Then the variation of (\ref{f1.7}) with
respect to {\it g}$^{\mu \nu }$ and $\phi $ gives, respectively, the field
equations

\begin{equation}
\begin{array}{r}
^{(3)}R_{\mu \nu }-\frac 12g_{\mu \nu }~^{(3)}R=-\frac{8\pi }\phi ~^{\left(
3\right) }T_{\mu \nu }-\text{\qquad \qquad \qquad \qquad \qquad \qquad } \\ 
-\frac \omega {\phi ^2}\left( \phi _{;\mu }\phi _{;\nu }-\frac 12g_{\mu \nu
}\phi _{;\rho }\phi ^{;\rho }\right) -\frac 1\phi \left( \phi _{;\mu ;\nu
}-g_{\mu \nu }\phi _{;\rho }^{;\rho }\right) ,
\end{array}
\label{f1.8}
\end{equation}

\begin{center}
\begin{eqnarray}
\phi _{;\mu }^{;\mu }=-\frac{8\pi }{3+2\omega }T,  \label{f1.9}
\end{eqnarray}
\end{center}

where $^{(3)}T_{\mu \nu }$ is the energy-momentum tensor of matter fields.

The system of equations (\ref{f1.8}), (\ref{f1.9}) have the vacuum solution 
{\it g}$^{\mu \nu }$ = $\eta ^{\mu \nu }$ and $\phi $ = const, where $\eta
^{\mu \nu }$ is the metric tensor of a flat space. Obviously, this solution
identical to Minkowski's space of general relativity and this solution fixes
the arbitrariness in the coordinate system. Notice that this explanation
viable in a more general theories when $f\left( \phi \right) \neq \phi $ and 
$\omega $($\phi $) $\neq $ const. It is necessary to remark that, according
to Mach, the inertial forces observed locally in accelerated laboratory may
be interpreted as gravitational effects having their origin in distant
matter accelerated relative to laboratory. Consequently, there must be
energy density of matter everywhere in machian universe. In completely empty
space there can not be no gravitational field at all; in this case it would
not be possible neither distribution of light, nor existence of scales and
hours \cite{c20}. In this case we must choose the solution {\it g}$^{\mu \nu
}$ = $0$ and $\phi $ = 0. On the other hand if we choose the case {\it g}$%
^{\mu \nu }$ = $\eta ^{\mu \nu }$ and $\phi $ = 0 then we came to
''Newtonian'' world.

The most appropriate coordinates in which to study the static spherically
symmetric field are isotropic. Hence, the metric will be assumed to have the
form

\begin{eqnarray}
ds^2=e^\beta \left( dr^2+r^2\left( d\theta ^2+\sin ^2\theta \,d\varphi
^2\right) \right) ,  \label{f1.10}
\end{eqnarray}

where, $\beta $ and $\phi $ will be assumed to be function of {\it r }only.
Expressing the line element in isotropic form equations (\ref{f1.8}), (\ref
{f1.9}) gives:

\begin{eqnarray}
\begin{array}{c}
\frac{\beta ^{\prime }\phi ^{\prime }}\phi -\frac{2\beta ^{\prime }}%
r-2\omega \left( \frac{\,\phi ^{\prime \,}}\phi \right) ^2-2\beta ^{\prime
\prime }-\frac{2\phi ^{\prime \prime }}\phi =0 \\ 
\\ 
\frac{3\beta ^{\prime }}r+\frac{\beta ^{\prime }\,^2}2+\,\beta ^{\prime
\prime }+\left( \frac 2r+\beta ^{\prime }\right) \frac{\phi ^{\prime }}\phi
=0 \\ 
\\ 
\frac 4r+\beta ^{\prime }+\frac{2\phi ^{\prime \prime }}{\phi ^{\prime }}=0
\end{array}
\label{f1.11}
\end{eqnarray}

The solution of equations (\ref{f1.11}) for the static sphericaly symmetric
configuration can be easily obtained. This gives

\begin{eqnarray}
\begin{array}{c}
\beta =-\ln \left( \frac{r^4\left( \frac{r+\mu }{r-\mu }\right) ^{\sqrt{%
\frac 8{2+\omega }}}}{\left( 2+\omega \right) \left( r^2-\mu ^2\right) ^2}%
\right) +c_1 \\ 
\\ 
\phi =c_2\left( \frac{r+\mu }{r-\mu }\right) ^{\sqrt{\frac 2{2+\omega }}}
\end{array}
,  \label{f1.12}
\end{eqnarray}

where $\mu ,$c$_1$ and c$_2$ arbitrary constant.

A point to be noted is that under choices the case $\omega $=0 the solution (%
\ref{f1.12}) becomes identical to Schwarzschild solution of general
relativity. It is well known that the exterior Schwarzschild solution of
general relativity does not incorporate Mach's principle. Moreover, the
anti-Machian character of the general relativity field equations also shows
up in solutions which allow for a curved spacetime even in the absence of
any matter \cite{c21}. The point is that the field equations admit curved
vacuum solutions and also solutions that are asymptotically flat. However, a
single mass point have inertia but in a consistent relativity theory there
cannot be inertia relative to ''space'' but in a Mach's philosophy the
inertia of a body gets determined by the presence of all other bodies in the
universe.

However, that despite the mathematical equivalence of the Schwarzschild and (%
\ref{f1.12}) with $\omega $=0 solutions, the fact that they come from
different physical theories makes them conceptually distinct. It is
surprising that in 3- dimensional temporal scalar tensor theories there is
not single matter particle solution. Thus in this sense the fourth dimension
(time) generates effective matter sources and in turn act to curve
3-dimensional space. In other words, existence of even a sole particle in
the space fills the entire universe with matter. Consequently, according to
the Machian hypothesis \cite{c25} there are the matter ''everywhere'' in
this universe.

\section{Conclusion}

There has been a lot of activity on describes our world as a 4-dimensional
surface (brane) world embedded in higher dimensional space-time (bulk) \cite
{c1}, \cite{c2}, \cite{c3}, \cite{c4}, \cite{c5}. This concept leads to a
great variety of specific models both in cosmological context and in the
description of local self-gravitating objects. Modifying the usual picture,
where the our Universe represent as 4-dimensional differentiable manifolds,
these our developments are based on the idea that matter fields could be
confined to 3-dimensional world, while time could live in a higher
dimensional space. So, like many authors, we try to describe N-dimensional
gravitational models with extra fields were obtained from some
multidimensional model by dimensional reduction based on decomposition of
the manifold \cite{c13}. Most of theories built on this idea represent the
4-dimensional space-time and its associated extra space by differentiable
manifolds hence such a gravity theory is equivalent to the Jordan,
Brans-Dicke theory with vanishing parameter $\omega $ and a potential term 
\cite{c16}, \cite{c23}. On the other hand after the conceptual foundations
of non-Euclidian geometry had been laid by Lobachevsky, Bolyai, Gauss and
the examinations on mechanics in non-Euclidean spaces had been clarified (at
the end of XIX centuries \cite{c26}, \cite{c27}), it was natural to consider
how Newton's law should fit into the new framework. In this scheme, in the
spirit of Newtonian theory, space and time are treated in completely
different ways.

We have presented here a new point of view to the General relativity theory.
In present article the possibility of description of 3-dimensional gravity
theories in terms of dimensional reduction is analyzed. By the way the idea
that the matter in three dimensions can be explained from a 4-dimensional
Riemannian manifold is a consequence of this point of view. This theory
represents some modification of general relativity can be interpreted as a
dynamical theory of evolution of 3- dimensional Riemannian geometry. A
related question in whether plausible brane physics can generically be
recovered and whether there are testable correction to Newtonian gravity. It
would be interesting to consider these question further.

If the brane world is real, one may find some evidences of
higher-dimensions. The {\it present} can be viewed as confined to a
time-brane, a 3-dimensional hyper-surface embedded in space-time with time
extra dimensions. It is necessary to remark that, the man feels the{\it \
present} (the thickness of brane) as 0.5 seconds. The field equations of
this theory formally have a structure similar to the Jordan, Brans-Dicke
theory in three dimension with the free parameter $\omega $ = 0. It is
expected on the general ground that dimensional reduction of a
multi-dimensional theory in the 3-dimensional scheme yields a temporal
scalar field theory. Consequently, these ideas may be continued by suggested
considering the more general case leading to the new theory with an
additional temporal scalar field. These models based on temporal scalars
have the theoretically appealing property that the same scalar field that
participates in the gravity sector is enhanced by potential. Introducing the
temporal scalar field a different actions (\ref{f1.7}) with different
choices of functions $\lambda $($\omega $), $f\left( \phi \right) $and $%
\omega $($\phi $) has been proposed.

Finally an important question with regards to three dimensional gravity
theories with temporal scalar field theories is how to choose reasonable
forms of the field equations and the energy-momentum tensor. An examination
based upon case of static spherically symmetric configuration with
restriction to the functions $\lambda $($\omega $) = 0, $f\left( \phi
\right) =\phi ~$and $\omega $($\phi $) = const., show that there is solution
of fields equations (\ref{f1.11}) identical to the Schwarzschild solution in
general relativity. However, the geometrical curvature induces matter in
three dimensions, and the extra dimensions (time) appear as the additional
effective matter source for gravity.

\end{document}